# Modeling of A Realistic DC Source in A CVSR

Mohammadali Hayerikhiyavi, Aleksandar Dimitrovski
Department of Electrical and Computer Engineering
University of Central Florida
Orlando, USA
Mohammad.ali.hayeri@knights.ucf.edu, ADimitrovski@ ucf.edu

*Abstract*—Continuously Variable Series Reactor (CVSR) can adjust the total reactance in an ac circuit using the saturation characteristic of the ferromagnetic core, shared by an ac and a dc winding. The bias magnetic flux produced by the dc winding can regulate the equivalent ac inductance in order to control power flow, damp oscillations, or limit fault currents. Gyrator-Capacitor approach is used to model the interface between the magnetic and the electric circuits. Two different dc control source models are considered: the usual ideal dc source and a realistic dc source composed of a power electronics-based converter and an ac voltage source. This paper investigates CVSR's behaviour in terms of induced voltages across the ac winding, flux densities (B) throughout the CVSR core, and power interchange with the dc control circuit during normal conditions in both cases.

*Index Terms*-- Continuously Variable Series Reactor (CVSR), Gyrator-Capacitor (G-C) model, magnetic amplifier.

## I. INTRODUCTION

Power systems operate close to their limits due to the continuously increasing demand for electric energy and lagging investment in the power system infrastructure. Contingencies in generation, transmission, demand volatility and other events may lead to blackouts, which is of primary concern to power system operators [1-2]. A significant number of these problems could be alleviated with an appropriate ac power flow control. Devices for power-flow control include phase-shifting transformers, switched shunt-capacitors/inductors, and different types of power electronics-based flexible ac transmission systems (FACTS) controllers [3]. However, these devices can either provide only a coarse control or have a very high cost. The saturable reactor technology does not suffer from these disadvantages and provides an alternative option for reliable and cost-effective power flow control.

Continuously Variable Series Reactor (CVSR) is a series-connected saturable reactor that has variable reactance within its design boundaries [4]. By changing a relatively small dc bias current, continuous and smooth control of large amount of power flow on the ac side can be achieved. Besides power flow control, other applications of CVSR include oscillation damping and fault current limiting [5, 6]. Inserting an additional impedance in the circuit in series with the ac winding can also damp oscillations or decrease fault currents.

The dc source for the CVSR is a low voltage and current power electronics-based converter. This is in contrast to the FACTS controllers that use components with high rated power, because they are also part of the main power system. Usually, in analysis, the dc current is supplied by an ideal dc source [7]. This paper considers a more realistic dc bias source and looks into the details of the power electronics-based converter.

A Gyrator-Capacitor (G-C) model is a better approach for modeling magnetic circuits in detailed analyses of power system devices. It links directly electrical and magnetic circuits for an integral analysis of the complex hybrid system in the power magnetic device. With this approach, the analogy between the magneto-motive force (MMF) and the voltage is preserved, but the electrical current is analogous to the rate of change of the magnetic flux (i.e., flux is analogous to electrical charge), while magnetic permeance (reciprocal reluctance) is analogous to capacitance. A more detailed model should also consider magnetic circuit nonlinearities (hysteresis and saturation) for more accurate analysis [8].

In this paper, a more detailed G-C model of the CVSR with a realistic dc bias source is analyzed. It also includes hysteresis and saturation for the nonlinear magnetic core. Two different models of the bias source in the control circuit are considered: i) an ideal current source, and ii) a more realistic source that includes power electronics-based converter. The models have been created in MATLAB/Simulink®, and the results from simulations are presented in the case study.

The following is the paper outline. The basic concept of the CVSR is briefly reviewed in Section II. The more detailed G-C approach in modeling magnetic circuits with nonlinearities, including core saturation and hysteresis, is discussed in Section III. Section IV presents simulation results and analysis of the CVSR for the two different models of the dc bias, and conclusions are summarized in Section V.

## II. CONTINUOUSLY VARIABLE SERIES REACTOR

In the CVSR shown in Figure 1, an ac winding is wound around the middle leg of a three-legged magnetic circuit, and connected in series with the ac circuit that delivers power from the source to the load. Typically, the center leg has an air gap to obtain the desired reactance in normal conditions, as well as to prevent core saturation for normal load currents and ac fluxes. Two dc windings connected in series and fed by a dc source are wound around the outer legs. The reactance of the CVSR is controlled by the dc current. It is at its highest value when the core operates in the linear region of the B-H curve (with no or very small dc), and at its lowest value when the core is entirely saturated (at large dc) [9]. As a result, the equivalent reactance of the ac circuit is controlled by the dc. This variable reactance can be used for power flow control, oscillation damping, and fault current limitation.

At all times, the fluxes passing through the outer legs are a combination of the ac and dc fluxes. They add in one and subtract in the other, as shown in Figure 1. The induced voltages in the right and left part of the dc winding are proportional to $\frac{d\Phi_{right}}{dt}$ and $\frac{d\Phi_{left}}{dt}$, respectively, and the induced voltage across the entire dc windings is $V_{bias} = V_{right} - V_{left}$.

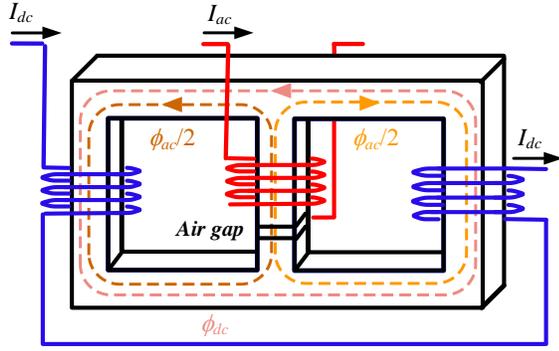

Fig. 1. Schematic of a CVSR

### III. GYRATOR-CAPACITOR MODEL

A common representation of magnetic circuits uses the electric circuit analogy [10], where resistors describe flux paths and voltage sources represent MMFs. Since magnetic circuits store energy, they are not adequately modeled by resistors which can only dissipate energy. The G-C model described in Figure 2 [11] retains this power equivalence.

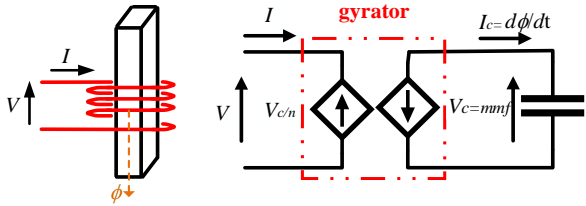

Fig. 2. A simple magnetic circuit and its equivalent gyrator-capacitor model

In the G-C approach, the analogy is between the MMF and the voltage, and the current and the rate-of-change of the magnetic flux $\frac{d\Phi}{dt}$, as expressed by (1) and (2) [12]:

$$V_c \equiv mmf \quad (1)$$

$$I_c = \frac{d\Phi}{dt} \quad (2)$$

These two expressions lead to an equivalent representation of magnetic permanences with capacitances. If a permeance is nonlinear, it is represented with a nonlinear capacitance. The permeance depends on the magnetic material and the geometric parameters of the magnetic circuit, and can be expressed approximately with (3):

$$\rho = \frac{\mu_r \mu_0 l}{A} \quad (3)$$

where: $\mu_0 = 4\pi \times 10^{-7}$ – magnetic permeability of free air, $\mu_r$ – relative magnetic permeability of the material, $A$ – cross-sectional area, $l$ – mean length of the path

A G-C model of the CVSR shown in Figure 1, as implemented in Simulink®, is shown in Figure 3. The nonlinear permeances representing nonlinear magnetic paths are modeled with variable capacitors, and a linear permeance models the air gap in the center leg. The windings are modeled with gyrators. Core hysteresis is modeled with resistors in the magnetic circuit [7].

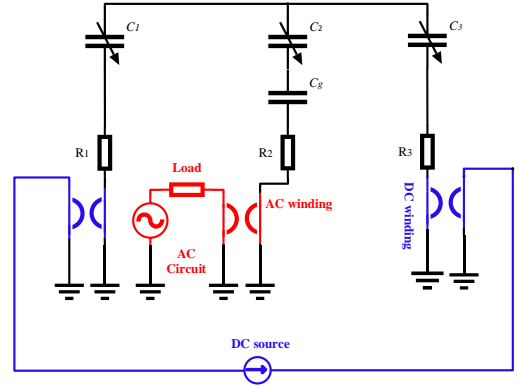

Fig. 3. Gyrator-Capacitor model in Simulink®

### IV. SIMULATIONS AND RESULTS

As mentioned above, the G-C model was implemented in Simulink® for dynamic simulations. This software was chosen for its component library available and compatibility with other modeling tools. All of the simulations were done on the CVSR model shown in Figure 3. The device and load parameters are as listed in Table I. The ferromagnetic core material is assumed to be M36 with details given in [12].

The dc control circuit is connected to the dc source via two gyrators that represent the two parts of the dc winding. The ac circuit contains a load and a voltage source. The gyrator in the middle represents the ac winding. It should be noted that the fringing effect of the air-gap flux, modeled for better accuracy, has been included inside the air gap permeance.

The main contribution of this paper is the introduction of a model of the power electronics-based converter in the dc circuit, powered from a low voltage ac source. To control the current applied to the dc winding, an H-bridge converter is composed of four IGBT (Insulated Gate Bipolar Transistor) switches, as shown in Figure 4 [13]. The converter's dc input voltage is provided by a diode-bridge rectifier powered from a low voltage ac outlet. A dc link capacitor filters the rectifier's output voltage before applying it to the converter's input. The output dc current applied to the control winding is continuously monitored and compared with its reference in a closed-loop control. The error between the measured and the reference current is measured, and appropriate gate switching signals are generated by the PI controller using pulse width modulation (PWM) with a corresponding switching frequency

(50 kHz used here). The closed-loop control system ensures that the output current follows the reference, which, in general, can be dc, ac, or ac with dc bias. The reference current can be changed dynamically according to the conditions. This flexibility is a significant feature that allows use in applications such as power flow control, oscillation damping and others. The PI controller is designed to provide acceptable overshoot, settling time, and ripple in a steady state. Each of the gains (proportional and integral) is calculated using the Zigler-Nicholes method to achieve desire overshoot and settling time.

The behavior of the CVSR with both sources is analyzed in terms of induced voltages across the windings and magnetic flux densities (B) in the core. Three different characteristic dc bias currents have been considered: 0 A, 5 A, and 30 A, for the different normal operating conditions of the device.

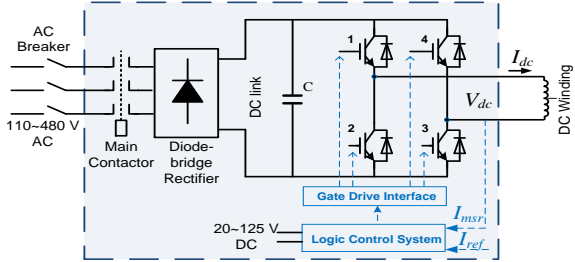

Fig 4. Power electronics-based dc converter and control system

Table I – CVSR Parameters

| Parameter | Description | Value |
|---|---|---|
| $l_m$ | mean length of the middle leg | 45.72 cm |
| $l_{out}$ | mean length of the outer legs | 86.36 cm |
| $h_{ag}$ | height of the air gap | 0.1780 cm |
| $A$ | cross-section area of the core | 0.0103 m² |
| $N_{dc}$ | number of turns in the dc winding | 30 |
| $N_{ac}$ | number of turns in the ac winding | 20 |
| $V$ | voltage source | 2.4 kV |
| $R$ | load resistance | 100 Ω |
| $L$ | load inductance | 130 mH |
| $B_{sat}$ | saturation point | 1.34 T |

### A. Ideal dc source

The flux densities and the terminal voltage (induced voltage across the ac winding) for the G-C model with 0 A dc bias are shown in Figures 5 and 6, respectively. The flux densities are purely sinusoidal and so is the terminal voltage. The ac flux is divided equally between the outer legs. Therefore, the induced voltages in the two parts of the dc windings are equal, opposite, and cancel each other.

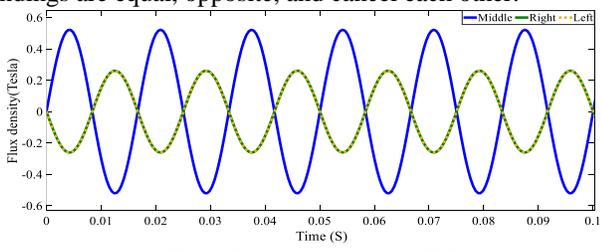

Fig. 5. Flux densities through CVSR ($I_{dc}$ =0A)

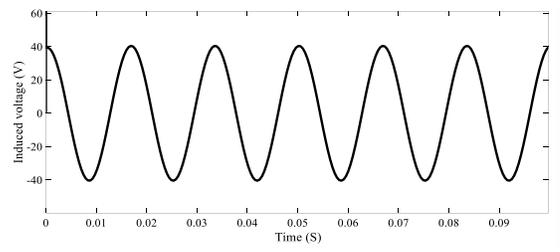

Fig. 6. Induced voltage across the ac winding ($I_{dc}$ = 0 A)

The current that passes through the ac winding is determined by the source voltage and the equivalent ac circuit impedance, which is a combination of the load impedance and the equivalent inductance of the CVSR in series. The latter is determined by the equivalent ac self-inductance of the CVSR given by (3):

$$L = N_{ac}^2/R_m \qquad (3)$$

where: $R_m$ – equivalent reluctance of the CVSR, and $N_{ac}$ – number of turns of the ac winding. Therefore, the current passes through the ac winding with 0 A dc has 21.2 RMS value.

The core flux densities for 5 A dc bias (a critical current), are shown in Figure 7. There is an offset in the outer flux densities and they are no longer identical. Also, they become distorted alternately when the corresponding leg enters into saturation for one half of the sinusoid. The induced voltage across the bias winding is no longer zero and has double the system frequency (Figure 8). The induced voltage across the ac winding starts to become distorted as well, as shown in Figure 9. Still, the current through the ac winding is almost the same as at 0 A dc.

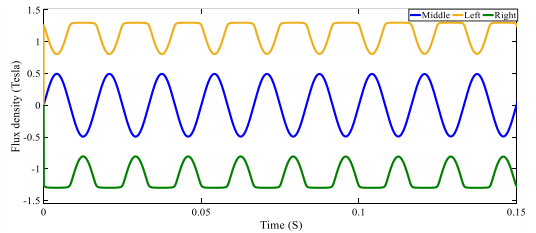

Fig7. Flux densities through CVSR ($I_{dc}$ =5A)

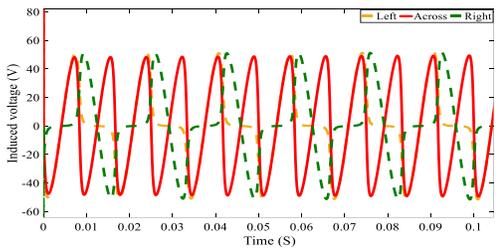

Fig8. Induced voltage across the dc windings ($I_{dc}$ = 5 A)

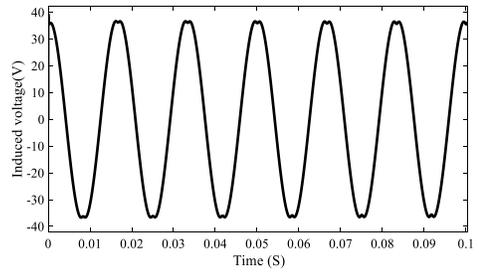

Fig. 9. Induced voltage across the ac winding ($I_{dc}$ = 5A)

Increasing the dc to a high value leads the core into complete saturation, as shown in Figure 10. The flux density through the center leg is minimal due to the very high reluctance of the saturated outer legs. Hence, the induced voltage across the ac winding is very small (Figure 11). For the same reason, the CVSR reactance is negligible, and the current in the ac winding is equal to the full load current in the circuit. Due to the full saturation of the outer legs, the induced voltage in the parts of the dc winding is the same, and the induced voltage across the winding is zero.

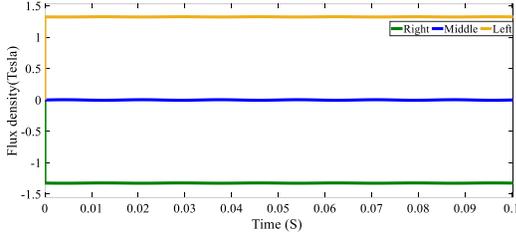

Fig 10. Flux densities through CVSR legs ($I_{dc}$ =30A)

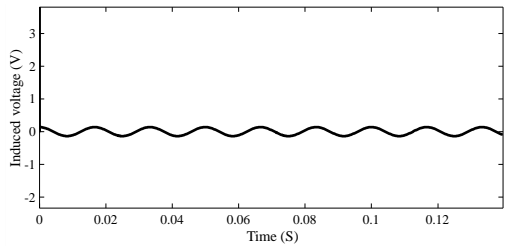

Fig. 11. Induced voltage across the ac winding ($I_{dc}$ = 30A)

### B. Dc source with power electronics-based converter

The flux densities at 0 A dc bias are shown in Figure 12. Unlike the current generated by the ideal dc source, the current that passes through the dc winding is not precisely zero. After the settling time, it exhibits a small ripple. It appears like there is a slight dc bias through the dc winding, as the flux densities in the outer legs are not identical and completely overlapped. In the first few milliseconds of the simulation there is a visible transient due to the transient response of the controller set to 0 A dc bias. Now, due to the ripple, the induced voltage across the dc winding is no longer zero. Figure 13 shows this with hysteresis, and Figure 14 without it (note that the x-axis scale in Figure 13&14 is zoomed in). The frequency is 50 kHz, that of the switching frequency. It should be noted that, since the bias flux circulates in the outer frame of the core, it does not affect the ac flux through the middle leg in the linear region. Therefore, the induced voltage across the ac winding and the current through it are not distorted. In comparison to the previous case (0 A ideal source) since the flux densities in the outer legs do not completely overlap, the induced voltage across dc windings will not be zero due to the hysteresis. Therefore, the power transferred into the dc windings is small but not zero (because the current passing through the winding is also not zero), as it was in the previous two cases.

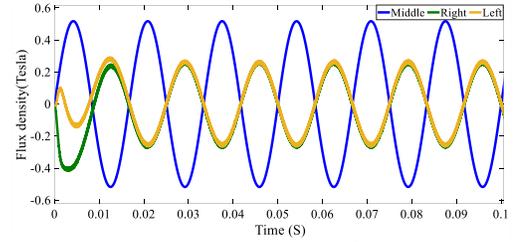

Fig12. Flux densities through CVSR ($I_{dc}$ =0A)

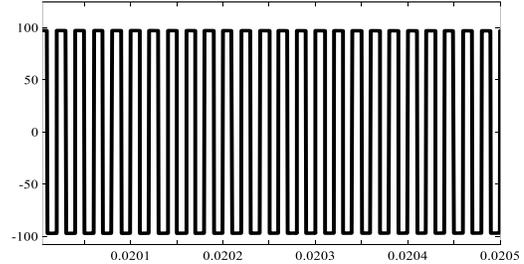

Fig13. Induced voltage across dc winding with hysteresis ($I_{dc}$ = 0 A)

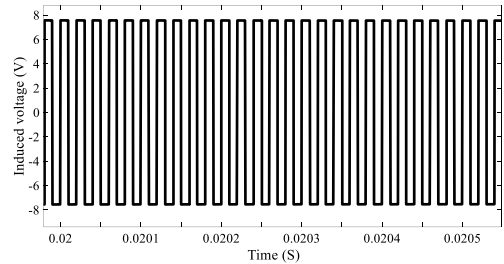

Fig14. Induced voltage across dc winding without hysteresis ($I_{dc}$ = 0 A)

Figure 15 shows the flux densities at the critical dc current. It is evident in this figure first cycle the transient due to the current overshoot from the controller. For the same reason, the dip in induced voltage across the ac winding in the first cycle is larger (Figure 16). The induced voltage across the dc winding and the power delivered to the dc winding are shown in Figures 17 and 18, respectively. The Figure 19 shows that the response of the proportional controller to provide dc current equals 5A passes through the DC winding to show the desired operation of the PI controller.

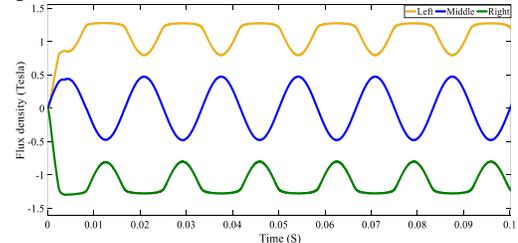

Fig. 15. Flux densities through CVSR legs ($I_{dc}$ =5A)

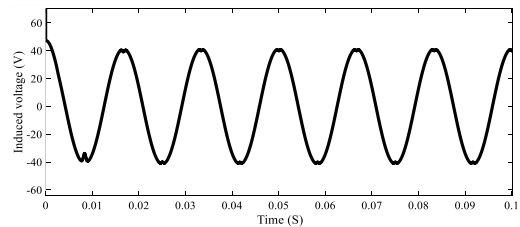

Fig16. Induced voltage across the ac windings ($I_{dc}$ = 5 A)

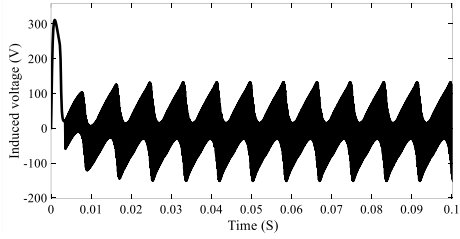

Fig17. Induced voltage across the dc winding ($I_{dc}$ = 5 A)

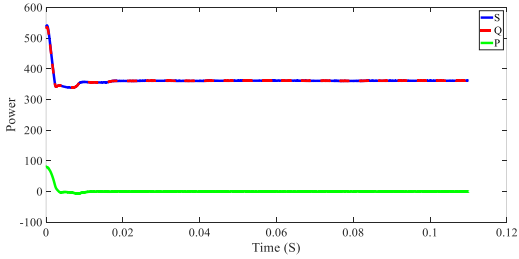

Fig. 18. Powers transferred to the dc winding ($I_{dc}$ = 5 A)

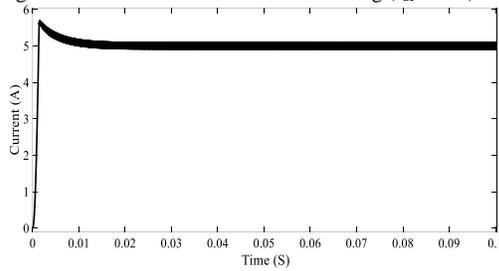

Fig. 19. output current of the controller supplying DC winding (DC=5A)

In Figures 20 and 21, the flux densities and the induced voltage across the dc winding are shown for the high dc bias current of 30 A, which drives the core into deep saturation. Again, the ac flux density through the center leg in the steady-state condition is small due to the high reluctance of the core. Therefore, the induced voltage across the ac winding is also small and so is the induced voltage across the dc winding as shown in Figure 22.

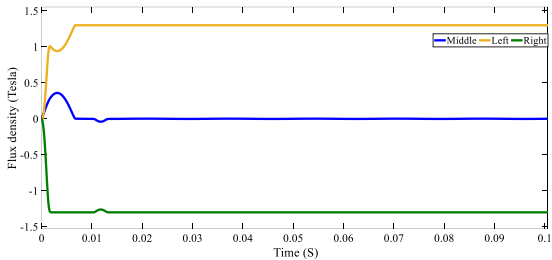

Fig20. Flux densities through CVSR legs ($I_{dc}$ =30A)

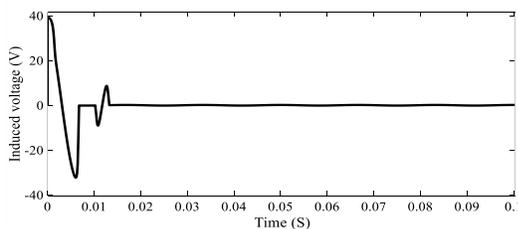

Fig21. Induced voltage across ac winding ($I_{dc}$ =30A)

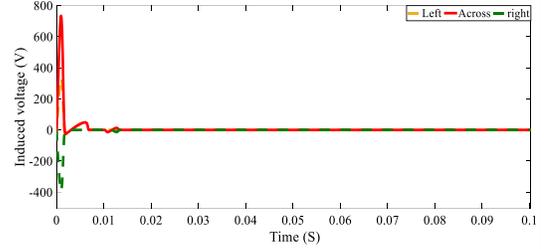

Fig22. Induced voltage across dc winding ($I_{dc}$ =30A)

## V. CONCLUSIONS

The paper presents an analysis of CVSR in different modes of operation, in terms of induced voltages across the windings and core flux densities. A more realistic, power electronics-based model of the bias dc source has been introduced to the basic model. A G-C modeling approach was used to model the complex magnetic-electric-electronic system. Results from the simulations of two different cases in steady state, one with an ideal dc source and the other with the realistic power electronic converter are presented and compared.

Future work will study a more accurate model of the device and its behavior during transient conditions.